\documentclass[pmlr]{jmlr}

\usepackage{microtype} 

\SetKwBlock{Loop}{Loop}{end}
\SetKw{Continue}{continue}
\SetKw{Break}{break}
\SetArgSty{textnormal}

\usepackage{longtable}

\usepackage{booktabs}
\usepackage[load-configurations=version-1]{siunitx} 

\usepackage[frozencache=true,cachedir=minted-cache]{minted}
\setminted[postgres]{
      frame=lines,
      framesep=3mm,
      baselinestretch=1.0,
      fontsize=\footnotesize,
      linenos
}
\renewcommand*\cite{\citep} 

\usepackage{enumitem}
\setlist{nolistsep}

\renewcommand*{\footnoteseptext}{. }

\jmlrvolume{1}
\jmlryear{2024}
\jmlrworkshop{Learning and Automata Workshop}
\jmlrpages{12}

\title[Database-assisted automata learning]{Database-assisted automata learning}

\author{\Name{Hielke Walinga} \Email{H.Walinga@student.tudelft.nl}\\
\Name{Robert Baumgartner} \Email{R.Baumgartner-1@tudelft.nl}\\
\Name{Sicco Verwer} \Email{S.E.Verwer@tudelft.nl}\\
\addr Software Technology Department, EEMCS, Delft University of Technology, Delft, the Netherlands
}

\begin{document}

\maketitle

\begin{abstract}
This paper presents DAALder (Database-Assisted Automata Learning, with Dutch suffix from \textit{leerder}), a new algorithm for learning state machines, or automata, specifically deterministic finite-state automata (DFA).
When learning state machines from log data originating from software systems,
the large amount of log data can pose a challenge.
Conventional state merging algorithms cannot efficiently deal with this,
as they require a large amount of memory.
To solve this, we utilized database technologies
to efficiently query a big trace dataset and construct a state machine from it,
as databases allow to save large amounts of data on disk while still being able to query it efficiently.
Building on research in both active learning and passive learning,
the proposed algorithm is a combination of the two.
It can quickly find a characteristic set of traces from a database using
heuristics from a state merging algorithm.
Experiments show that our algorithm has similar performance to
conventional state merging algorithms on large datasets,
but requires far less memory.
\end{abstract}

\section{Introduction}\label{sec:introduction}

Automata can be helpful models for analyzing software systems.
For example, state machines have helped to analyze software for
web protocols \citep{bertolinoAutomaticSynthesisBehavior2009},
X11 programs~\cite{ammonsMiningSpecifications2002},
and SSH~\cite{fiterau-brosteanModelLearningModel2017}.
They have also helped to test software systems \citep{hagererEfficientRegressionTesting2001},
to check software protocols~\cite{choMACEModelinferenceAssistedConcolic2011,comparettiProspexProtocolSpecification2009},
or for reverse engineering~\cite{antunesReverseEngineeringProtocols2011,cuiDiscovererAutomaticProtocol2007}.

Finding a corresponding state machine for a software system is a grammatical inference problem~\cite{delahigueraGrammaticalInferenceLearning2010}.
Techniques for inferring the state machine can be divided into two categories.
One is \textit{active learning}, algorithms that construct the state machine by providing clever inputs to the system and, during that process, learn the state machine~\cite{angluinLearningRegularSets1987}.
The other is \textit{passive learning}, algorithms that process a large amount of inputs and outputs,
and infer from that data the underlying states~\cite{cookDiscoveringModelsSoftware1998}.

A problem for \textit{passive learning} is that all input data has to be present at the start.
This leads to a very large memory footprint at the start because usually, all data present is used to construct an observation tree (a PTA, prefix tree acceptor).
Solutions for this problem could be sampling the initial data, or learning the automata using a streaming approach~\cite{balleBootstrappingLearningPDFA2012,balleAdaptivelyLearningProbabilistic2014,baumgartnerLearningStateMachines2022,baumgartnerLearningStateMachines2023,schmidtOnlineInductionProbabilistic2014}.

In this paper we present the algorithm DAALder that is a combination of active and passive learning.
Instead of loading all the available data in memory,
the data is saved on disk in a database.
Decades of engineering hours have gone into developing databases that allow for efficient querying of its data.
It is also often the original data source.
Especially when learning from software logs, the logging information is frequently stored in a large distributed database, for example, the very popular Splunk database\footnote{\url{https://www.splunk.com/}}.
Here, we develop a type of active learning algorithm that, instead of asking membership and equivalence queries, asks database queries to iteratively construct a state machine.

A key benefit of our approach is that membership queries in traditional active learning ask for a very specific set of traces. It is likely that several of the asked traces are not present in the collected data. In this case, it is unclear how to proceed.
There is research on active learning algorithms in a setting where not all queries can be answered.
In these settings, the teacher, the component that answers the query,
is referred to as an inexperienced teacher~\cite{leuckerLearningMinimalDeterministic2012}
or incomplete teacher~\cite{moellerAutomataLearningIncomplete2023}.
These approaches rely on SAT solving~\cite{grinchteinLearningFiniteStateMachines2006,moellerAutomataLearningIncomplete2023},
use containment queries~\cite{chenLearningMinimalSeparating2009},
or loosen L* with weakened concepts~\cite{grinchteinInferringNetworkInvariants2006}.
All these approaches have the CSP (constraint-satisfaction problem)~\cite{biermannSynthesisFiniteStateMachines1972} as a basis for their own approach~\cite{leuckerLearningMinimalDeterministic2012}.
Relying on SAT solvers, however, can take a lot of time~\cite{moellerAutomataLearningIncomplete2023}.

Intuitively, our approach avoids asking for this specific set of traces, which eventually forms a characteristic sample~\cite{delahigueraGrammaticalInferenceLearning2010}. For passive automaton learning algorithms, the presence of such a sample in the data implies that greedy algorithms such as RPNI~\cite{oncinaIdentifyingRegularLanguages1992} are guaranteed to find a model for (converges to) the target language (that is assumed to have generated the data). There exist many possible characteristic samples, but traditional active learning only constructs a specific one. Our main goal is to use a combination of active and passive learning such that the algorithm converges to the target when any characteristic sample is present in the database, while observing only a fraction of the data. Our contributions are:

\begin{itemize}
\setlength\itemsep{0em}
\item A new algorithm, named DAALder, to learn a state machine from a database of traces.
\item An implementation\footnote{Available after acceptance of the paper in \url{https://github.com/tudelft-cda-lab/FlexFringe}} of this algorithm in the state merging software package FlexFringe \cite{verwerFlexfringePassiveAutomaton2017,verwerFlexFringeModelingSoftware2022}.
\item Results on the performance of DAALder compared with a conventional EDSM (Evidence-Driven State Merging) algorithm on datasets of increasing size.
\end{itemize}


%

\section{Algorithm Description}\label{sec:algorithm-description}


\subsection{Queries}\label{sec:prefix-query}

For conventional active learning, the membership query and the equivalence query exist. In practice, it is not feasible to perform an equivalence query exactly. Often, a randomized search is employed, which is also done for this implementation. (See Appendix \ref{apd:randomized}.)
Membership queries are used to grow the model.
A membership query returns given an input sequence (a trace) its output symbol.
However, using a database enables us the freedom to design queries
that are more powerful than membership queries, yet easy to compute.

The novel query used in this algorithm is the prefix query.
This query is noted as \textsc{PrefixQuery($t$, $n$, $k$)} with $t$ being the prefix trace, $n$ being the max size of the returning traces, and $k$ the amount of traces to return.
This query returns up to a number of $k$ traces alongside their output symbol that has $t$ as their prefix.
Parameter $n$ can limit the result by their length to explore only a specific depth.
Implementation details in Appendix \ref{apd:prefix}.


\subsection{Algorithm: DAALder}

DAALder, the algorithm presented here, is a form of a state merging algorithm in the red-blue-framework~\cite{langResultsAbbadingoOne1998}.
However, instead of constructing a PTA directly from all the data available,
it is constructed iteratively using the at that moment available traces from the complete data that the algorithm deemed informative.

$L^{\#}$~\cite{vaandragerNewApproachActive2022,vaandragerNewApproachActive2022a} was a large inspiration for DAALder.
Just like the $L^{\#}$ algorithm, DAALder works directly on a PTA data structure (or observation tree).
This gives rise to an efficient way to search for new states and generate hypotheses.

\subsubsection{Outline}
\begin{figure}[ht]
\centering
\includegraphics[scale=1]{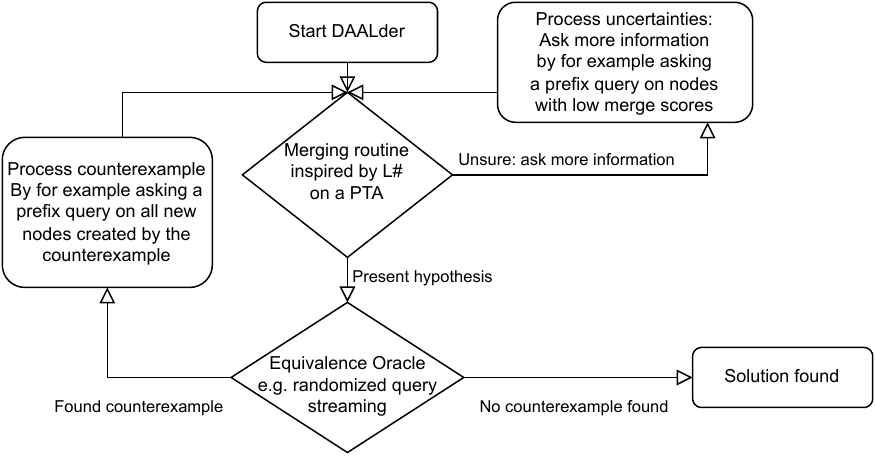}
\caption{Graphical representation of DAALder}
\label{fig:DAALder_graphical}
\end{figure}

Pseudocode and an outline of DAALder are found in algorithm \ref{alg:daalder} and in figure \ref{fig:DAALder_graphical}, respectively.

DAALder uses a state merging routine similar to $L^{\#}$.
It performs multiple rounds of state merging, each time with more data.
Occasionally, it will present a hypothesis to the oracle in the hopes of providing the correct hypothesis.

From EDSM, we know that information on the traces that pass through nodes can be used to determine whether two nodes are equivalent and, thus, can be merged.
Given some scoring heuristics, merges can be compared, and when a merge has the largest score, the merge is likely merging equivalent states~\cite{langResultsAbbadingoOne1998}.

DAALder uses these exact scoring heuristics to, next to selecting merges, also guide what queries it should ask.
If during merging, for the same blue node, merges with a similar score are encountered, those merges are put on hold.
Instead of selecting one of these uncertain merges,
it asks for more data for the nodes involved in them and thus providing the inconclusive nodes with more information. (in \textsc{DAALderProcessUnidentified})

DAALder will then ask prefix queries for the nodes involved in these uncertain merges.
Traces returned by these queries will contain more information for the scoring heuristics when calculating scores again for those nodes.
Traces gained by this are added to the observation tree whenever a new merging routine is started.
This new data can then be used to be more certain about merges to perform in the new round that previously were doubtful.

After merging, DAALder can either present its hypothesis to the equivalence oracle
or reset the tree and start a new round of merging.
If there had been a lot of new information,
it is wise just to reset the tree, because a lot of new information might imply the current hypothesis is wrong and the oracle takes much time.
If presented to the oracle, the oracle can answer "yes" or present a counterexample to be processed, and the tree is also reset.

In the end, when it finds a sufficiently correct state machine,
it has gathered a subset of traces that contain enough information to construct a correct hypothesis.
This is a so-called characteristic set of traces~\cite{delahigueraGrammaticalInferenceLearning2010}, or tell-tale set~\cite{angluinInductiveInferenceFormal1980}. 

DAALder is thus a combination of active and passive learning.
It performs a search strategy to find new information based on the state-of-the-art active learning approach from $L^{\#}$
and the highly flexible state-of-the-art implementation of passive learning state merging algorithms from FlexFringe.
This also means that all variations and hyperparameters on state merging available in FlexFringe are also directly available to DAALder.

\noindent\begin{minipage}{\textwidth}
\renewcommand*{\footnoteseptext}{ }
\renewcommand\footnoterule{}  
\begin{algorithm2e}[H]
  \caption{DAALder algorithm}\label{alg:daalder}\footnotetext{\textsc{DAALProcessUnidentified} and \textsc{ProcessCounterExample} add new traces to the observation tree using the \textsc{PrefixQuery}.
In our implementation \textsc{DAALProcessUnidentified} asks for more information on nodes when the highest merge score is less than twice the second-largest merge score.}
\KwData{A set of traces with an output label}
\KwResult{A hypothesis $\mathcal{H}$ consistent with the provided traces}
\textbf{Initialize:} $\mathcal{T}$ a PTA (observation tree) with red core $S$ and blue fringe $F$ \\
\textbf{Initialize:} $S$ red core nodes, initially with a root node \\
\textbf{Initialize:} $F$ blue fringe nodes, initialize using some random traces \\
\Loop{
  \texttt{state\_isolated} $\gets false$ \\
  \texttt{unidentified} $\gets \varnothing$ \textit{(The blue states without merge candidate)}\\
  \texttt{all\_merges\_todo} $\gets \varnothing$ \textit{(Merges to do, to make the hypothesis state machine)} \\
  \ForEach{$p \in F$}{
    \texttt{possible\_merges} $\gets 0$ \\
    \texttt{merge\_candidate} $\gets\ ?$ \\
    \ForEach{$q \in S$} {
      \If{\textsc{CheckConsistency($p$, $q$)}} {
        \texttt{possible\_merges} $\gets \texttt{possible\_merges} + 1$ \\
        \texttt{merge\_candidate} $\gets q$
      }
    }
    \If{\texttt{possible\_merges} $= 1$}{
      \texttt{all\_merges\_todo} $add (p, \texttt{merge\_candidate})$ \textit{(identified node)}
    }
    \If{\texttt{possible\_merges} $= 0$}{
      \textsc{DAALPromote}($p$) \\
      \texttt{state\_isolated} $\gets true$
    }
    \If{\texttt{possible\_merges} $ > 1$}{
      \texttt{unidentified} $add\ p$ \textit{(Investigate these states more)}
    }
  }
  \texttt{(explore\_more, merges\_todo)} $\gets$ \textsc{DAALProcessUnidentified}(\texttt{unidentified}) \\
  \texttt{all\_merges\_todo} $\gets$ \texttt{all\_merges\_todo} $\cup$ \texttt{merges\_todo} \\
  \If{\texttt{explore\_more $or$ \texttt{state\_isolated}}}{
    \Continue \textit{(Investigate the added traces and/or a new fringe/frontier)}
  }
  $\mathcal{H} \gets \textsc{BuildHypothesis}(\mathcal{T},\ \texttt{all\_merges\_todo})$ \\
  $\sigma \gets \textsc{EquivalenceOracle}(\mathcal{H})$ \\
  \If{$\sigma$} {
    \textsc{ProcessCounterExample}($\mathcal{T}$, $\sigma$) \\
    \Continue \textit{(Did not found the solution yet)}
  }
  \Return $\mathcal{H}$ \textit{(Solution found!)}
}
\end{algorithm2e}
\end{minipage}

\subsubsection{Explore vs exploit}

Besides the choice of state merging heuristics, DAALder provides another way to tweak its inner workings.
This is the \textsc{DAALProcessUnidentified} subroutine.
Here, a choice must be made between adding more data to combat uncertainty or being more daring and asking the equivalence oracle earlier.
This is a choice between exploration (by asking for more data) and exploitation of the current data (by asking the equivalence oracle).
Asking the equivalence oracle could be a time-consuming process,
but so could be asking for more data, in addition to the memory required to add the data to the observation tree. 


\section{Algorithm Evaluation}\label{sec:algorithm-evaluation}
\subsection{Data generation}

Artificial trace data has been generated by a modified version of \texttt{FSM-learning} from \citet{giantamidisLearningMooreMachines2021}
available at \url{https://github.com/hwalinga/FSM-learning}.
The program is modified to use JSON serializing instead, as its binary format did not work on Linux.
It is also modified to use a streaming approach to output data instead of writing it at the end of the execution of the program to reduce memory requirements.

Artificial state machines were generated with \texttt{gen\_moore.py}.
The input and output alphabet were both set to a size of two.
The number of states was set to two hundred.
\citet{giantamidisLearningMooreMachines2021} took inspiration from \citep{langResultsAbbadingoOne1998} for this program.

The \texttt{generate} program was used to generate traces.
The program generates the traces using a random walk and maintains frequencies of states visited before which it uses to calculate probabilities of states to visit next during its random walk.

Every single dataset is one experiment and had its own randomized state machine.
The sizes for the datasets varied from 625 unique traces to 40.960.000 unique traces for the train data.
The lengths are between 2 and 64.
The test set had a fixed amount of traces of 1.000.000.

\subsection{Experiments}

The DAALder algorithm is compared with the FlexFringe implementation of EDSM (Evidence Driven State Merging).
For configuration see Appendix \ref{apd:edsm}.
EDSM is run with low memory (capped at 16GB with cgroups) and high memory (70GB).
These tests show EDSM's response to using swap memory.
To test the exploration-vs-exploitation balance for DAALder, its hyperparameter $k$
(see section \ref{sec:prefix-query})
was set at 10, 100, and 1000.

Memory usage in the experiments was measured as virtual memory.
This includes the swap memory.
The memory was measured using \texttt{ps -U postgres -o vsz}
while running the algorithms under the \textit{postgres} user as well.
This way, we also include any memory allocated by the database engine itself.

\section{Results}\label{sec:results}

\begin{figure}[htbp]
\begin{minipage}[b]{0.50\linewidth}
\centering
\includegraphics[width=\textwidth]{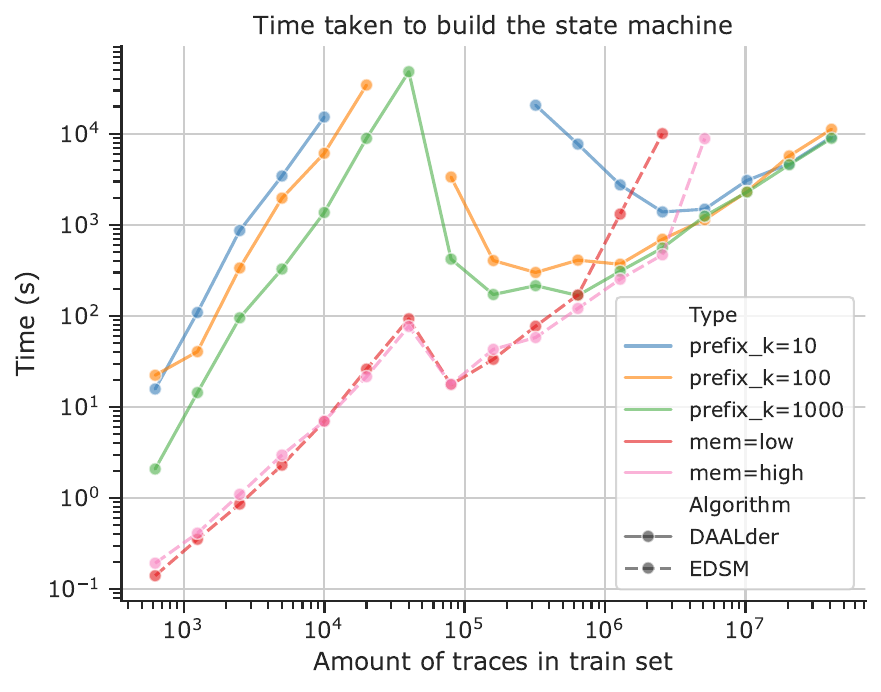}
\end{minipage}
\begin{minipage}[b]{0.50\linewidth}
\centering
\includegraphics[width=\textwidth]{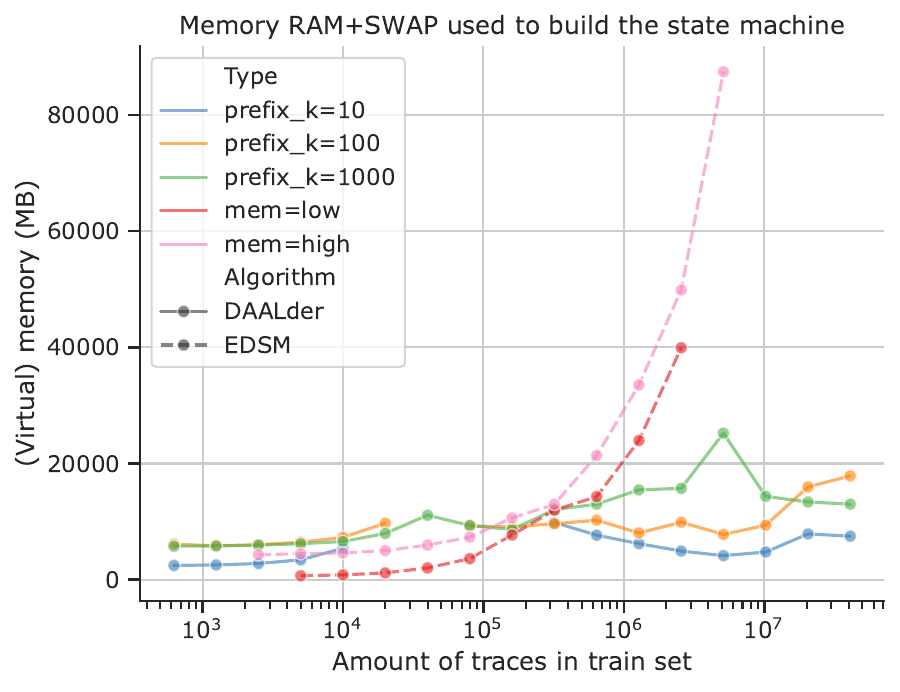}
\end{minipage}
\vspace*{-10mm}
\caption{Time and memory usage for EDSM vs DAALder}
\label{fig:timemem}
\end{figure}

\begin{figure}[htbp]
\begin{minipage}[b]{0.50\linewidth}
\centering
\includegraphics[width=\textwidth]{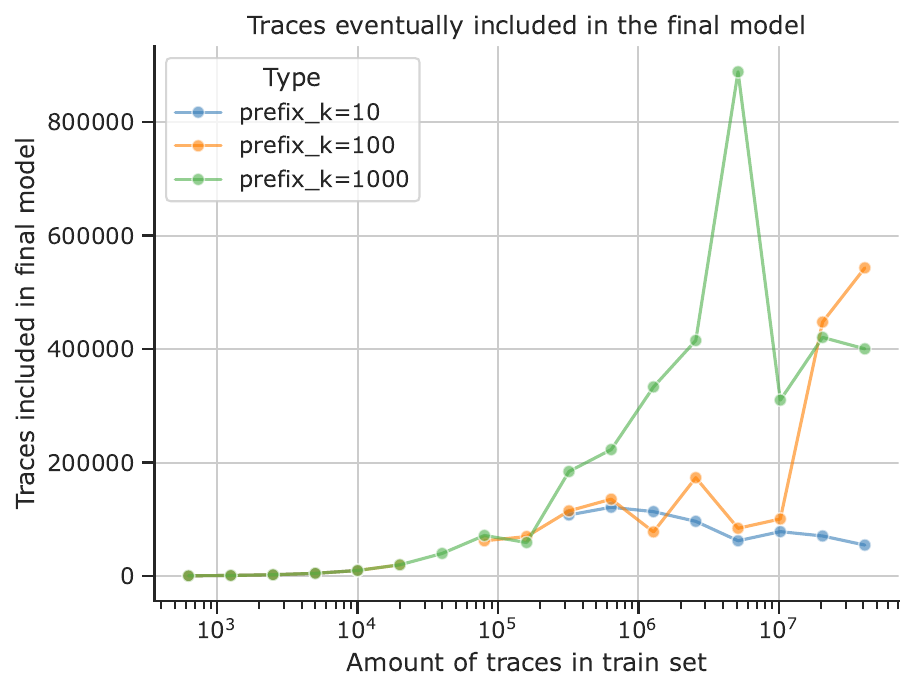}
\end{minipage}
\begin{minipage}[b]{0.50\linewidth}
\centering
\includegraphics[width=\textwidth]{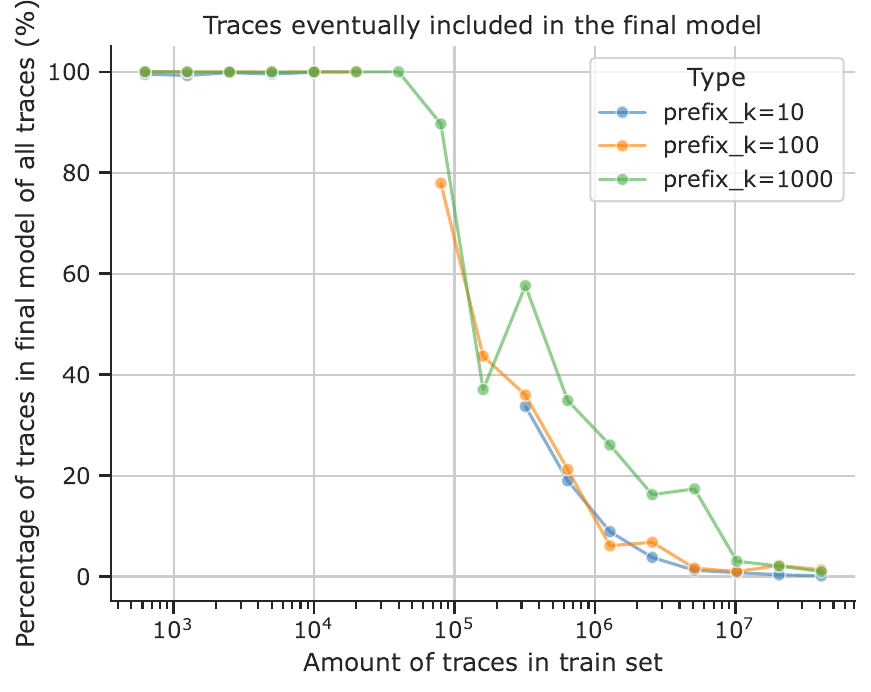}
\end{minipage}
\vspace*{-10mm}
\caption{Amount of traces included in the final model for DAALder}
\label{fig:traces}
\end{figure}

Figure \ref{fig:timemem} and \ref{fig:traces} show the results of the experiments.
Figure \ref{fig:timemem} depicts the relationships between the amount of traces versus the time the algorithm took to finish and the memory used by the algorithm,
Figure \ref{fig:traces} depicts the traces eventually included in the model (for DAALder)
and the percentage that this is of all the traces in the dataset.

Both EDSM and DAALder with the different parameters scored equally well on accuracy.
The algorithms scored perfectly for 80k traces or higher,
and for the datasets with 40k and lower, they scored just as good as guessing (50\%).

Some data is missing because for EDSM the biggest datasets ran out of memory (even using swap), trying to allocate more than 100 GB.
For DAALder data is missing around the 40k data points mark because running it took over 10+ hours and we choose to kill the program.
There are also no data points for memory reading for EDSM for the smallest datasets, as the program was too quickly done for a memory measurement to finish.

\section{Discussion}

\subsection{Solvability point}

As remarked earlier, there was a clear tipping point in the experiments based on the accuracy scores.
After the 40k data mark, all algorithms scored 100\% on accuracy; before that, they were just as good as just guessing the output class.
This means that the sampled data is after this point now big enough to contain a characteristic set~\cite{delahigueraGrammaticalInferenceLearning2010}.
This amount of data seems to be the solvability tipping point on these kinds of datasets. 

After the solvability point, the EDSM algorithm's runtime drops briefly.
It can now make the correct automaton, whereas before, it was heavily overfitted on the sparse data it had to work with, reporting a hypothesis with many more states than necessary.
It can also be seen in the log files that there is no ambiguity in merges to be made anymore.
That is, after the solvability point, there is always only one possible merge.

Around the solvability tipping point, DAALder performs the worst in terms of runtime,
also far worse than EDSM.
It took so long to finish that it had to be shut down preliminary.
As there is not enough data yet to arrive at a solution,
DAALder iteratively keeps including more data until (almost) all data is included.
This takes a lot of time.
This can also be seen in figure \ref{fig:timemem}, where there is a clear peak around this point at the 40.000 traces mark.

\subsection{Memory usage}

When EDSM runs out of RAM memory, it resorts to swap.
This behavior is very visible for the large datasets between the low and the high memory setups.
When EDSM had to resort to swapping, its runtime increased more steeply.
The slope for this increasing runtime was also significantly more than for DAALder for the largest datasets.
However, it seems that EDSM would perform on par with DAALder if given enough RAM.

The memory usage of EDSM increases exponentially.
So much so that not all experiments could be completed for EDSM.
For DAALder, the memory usage remained equal at just a few GBs.
This can be explained by the fact that DAALder ensures that just enough traces are included to create a correct model.
In fact, when $k$ was set to 10, the by DAALder selected set of traces was around the same size as the sample size around the solvability tipping point.
Thus, if explore settings are not too aggressive, DAALder is indeed able to select a set just large enough of the most informative traces to represent the state machine.

\subsection{Finding a characteristic set}

It is important to note that these tests were run on data from a statistical uniform sampler.
The effect of this is that for a random sample, the minimum size that contains a characteristic set is much smaller,
than if the data was generated with a non-uniform sampler.
If some transitions of the automaton are only described by a set of rare traces, these are less likely to be found in random samples of this data.
For those datasets, sampling data would not be a viable strategy.

For our uniform sampled data, DAALder used a set of traces that is somewhat equal in size to the sample size at a solvability tipping point.
However, for a different dataset with traces that traverse its state machine non-uniformly, we expect that DAALder is able to find a much smaller characteristic dataset,
than the sample size that would be required if it was randomly sampled.

\subsection{Effect of $k$}

Increasing $k$ (including more traces for each prefix query) for DAALder had the effect of also ending up with more traces in the end (and thus also more RAM).
For the sparse datasets, the runtime decreased with increasing $k$.
However, when the datasets grow larger, and sparsity decreases, the runtimes for different values of $k$ are very close to each other.
It seems that if sparsity decreases, a smaller set of traces from a prefix query is just as well able to distinguish a state as a larger set of traces from a prefix query.

\section{Conclusions and Future Work}\label{sec:conclusions-and-future-work}

This paper presents DAALder, a new algorithm combining previous work on active learning and passive learning.
It shows that this algorithm is faster than conventional state merging algorithms,
when these algorithms do not have enough RAM memory for very large datasets.

Future work requires checking how well DAALder performs on different kinds of datasets and state machines. It will focus on researching different approaches to the explore-exploit balance,
and finding what works best on these different datasets, and eventually learning from real-life datasets.

What is also of interest, is if we can increase our toolbox of queries.
Designing different kinds of queries that are optimized by different database technologies,
opens up more possible state machine learning algorithms that operate on trace databases.

One of these is to perform an equivalence query directly on the database.
Although regular expressions can be executed on databases, these implementations do not scale.
Furthermore, a consistency test can also directly be evaluated on the database for two states by comparing the prefix queries of the two access traces with each other.

Finally, to mitigate short-comings not a single strategy will suffice.
Ultimately, the holy grail is not one strategy, but an algorithm that can change strategies depending on the data it is dealing with.
An algorithm that can learn the best way to learn during its learning process.

\newpage
\appendix
\section{Implementation details}\label{apd:implementation}

\subsection{Randomized query}\label{apd:randomized}

Instead of checking a random primary key one by one, this can be done much more efficiently by using a pseudo-random streaming approach using libpqxx based on the PostGreSQL COPY
protocol. (\url{https://web.archive.org/web/20231212092242/https://pqxx.org/development/libpqxx/##2023-01-12-at-last-faster-than-c})

\subsection{Prefix query}\label{apd:prefix}

The prefix query is implemented as follows:

\begin{listing}[!ht]
  \begin{minted}{postgres}
SELECT * FROM table WHERE trace LIKE :'t' || '%' AND LENGTH(trace) < :n LIMIT :k;
\end{minted}
\vspace*{-5mm}
\caption{Select $k$ traces that start with $t$ and are shorter than $n$.}
\end{listing}

The prefix query can be sped up using the SP-GiST indexing, which is a similar data structure to a PTA, but optimized for disk usage~\cite{eltabakhSpacePartitioningTreesPostgreSQL2006}.
It should be noted that which $k$ traces are returned, is dependent on the database implementation and ordering of traces in the index structures.
The value of $n$ is not set for the experiments, so there is no additional filtering on the lenghts of the traces.

\section{EDSM parameters}\label{apd:edsm}

EDSM is run with parameters defined in \texttt{ini/edsm.ini}.
This specifies that there are no sink nodes and there is no lower bound.
In addition, only the blue node that currently has the most amount of traces following it, is considered for a merge, or a promotion to a red node if no merges are available to it.

\section{Setup}

The tests are run on a Dell PowerEdge R710 with an Intel Xeon CPU E5620, Seagate ST3146356SS, and Hynix HMT151R7BFR4C-H9 memory sticks.
Memory speed was measured with \texttt{sysbench} as well as the hard disk read speed with mode \texttt{rndrd}. The CPU speed was measured under \texttt{stress} from \texttt{/proc/cpuinfo}.
The performance is a memory speed of 4100 MiB/s, an HDD random read speed of 2.7 MiB/s, and a CPU speed of 2527 MHz.
The computer ran with Ubuntu 22 and postgresql 14.11, libpqxx 7.9.0, and clang 16.0.6.

\bibliography{natbib.bib}

\begin{thebibliography}{31}
\providecommand{\natexlab}[1]{#1}
\providecommand{\url}[1]{\texttt{#1}}
\expandafter\ifx\csname urlstyle\endcsname\relax
  \providecommand{\doi}[1]{doi: #1}\else
  \providecommand{\doi}{doi: \begingroup \urlstyle{rm}\Url}\fi

\bibitem[Ammons et~al.(2002)Ammons, Bod{\'i}k, and
  Larus]{ammonsMiningSpecifications2002}
Glenn Ammons, Rastislav Bod{\'i}k, and James~R. Larus.
\newblock Mining specifications.
\newblock \emph{ACM SIGPLAN Notices}, 37\penalty0 (1):\penalty0 4--16, January
  2002.
\newblock ISSN 0362-1340.
\newblock \doi{10.1145/565816.503275}.
\newblock URL \url{https://dl.acm.org/doi/10.1145/565816.503275}.

\bibitem[Angluin(1980)]{angluinInductiveInferenceFormal1980}
Dana Angluin.
\newblock Inductive inference of formal languages from positive data.
\newblock \emph{Information and Control}, 45\penalty0 (2):\penalty0 117--135,
  May 1980.
\newblock ISSN 0019-9958.
\newblock \doi{10.1016/S0019-9958(80)90285-5}.
\newblock URL
  \url{https://www.sciencedirect.com/science/article/pii/S0019995880902855}.

\bibitem[Angluin(1987)]{angluinLearningRegularSets1987}
Dana Angluin.
\newblock Learning regular sets from queries and counterexamples.
\newblock \emph{Information and Computation}, 75\penalty0 (2):\penalty0
  87--106, November 1987.
\newblock ISSN 08905401.
\newblock \doi{10.1016/0890-5401(87)90052-6}.
\newblock URL
  \url{https://linkinghub.elsevier.com/retrieve/pii/0890540187900526}.

\bibitem[Antunes et~al.(2011)Antunes, Neves, and
  Verissimo]{antunesReverseEngineeringProtocols2011}
Joao Antunes, Nuno Neves, and Paulo Verissimo.
\newblock Reverse {{Engineering}} of {{Protocols}} from {{Network Traces}}.
\newblock In \emph{2011 18th {{Working Conference}} on {{Reverse
  Engineering}}}, pages 169--178, Limerick, Ireland, October 2011. IEEE.
\newblock ISBN 978-1-4577-1948-6.
\newblock \doi{10.1109/WCRE.2011.28}.
\newblock URL \url{http://ieeexplore.ieee.org/document/6079839/}.

\bibitem[Balle et~al.(2012)Balle, Castro, and
  Gavald{\`a}]{balleBootstrappingLearningPDFA2012}
Borja Balle, Jorge Castro, and Ricard Gavald{\`a}.
\newblock Bootstrapping and {{Learning PDFA}} in {{Data Streams}}.
\newblock In \emph{Proceedings of the {{Eleventh International Conference}} on
  {{Grammatical Inference}}}, pages 34--48. PMLR, August 2012.
\newblock URL \url{https://proceedings.mlr.press/v21/balle12a.html}.

\bibitem[Balle et~al.(2014)Balle, Castro, and
  Gavald{\`a}]{balleAdaptivelyLearningProbabilistic2014}
Borja Balle, Jorge Castro, and Ricard Gavald{\`a}.
\newblock Adaptively learning probabilistic deterministic automata from data
  streams.
\newblock \emph{Machine Learning}, 96\penalty0 (1):\penalty0 99--127, July
  2014.
\newblock ISSN 1573-0565.
\newblock \doi{10.1007/s10994-013-5408-x}.
\newblock URL \url{https://doi.org/10.1007/s10994-013-5408-x}.

\bibitem[Baumgartner and Verwer(2022)]{baumgartnerLearningStateMachines2022}
Robert Baumgartner and Sicco Verwer.
\newblock Learning state machines via efficient hashing of future traces, July
  2022.
\newblock URL \url{http://arxiv.org/abs/2207.01516}.

\bibitem[Baumgartner and Verwer(2023)]{baumgartnerLearningStateMachines2023}
Robert Baumgartner and Sicco Verwer.
\newblock Learning state machines from data streams: {{A}} generic strategy and
  an improved heuristic.
\newblock In \emph{Proceedings of 16th Edition of the {{International
  Conference}} on {{Grammatical Inference}}}, pages 117--141. PMLR, July 2023.
\newblock URL \url{https://proceedings.mlr.press/v217/baumgartner23a.html}.

\bibitem[Bertolino et~al.(2009)Bertolino, Inverardi, Pelliccione, and
  Tivoli]{bertolinoAutomaticSynthesisBehavior2009}
Antonia Bertolino, Paola Inverardi, Patrizio Pelliccione, and Massimo Tivoli.
\newblock Automatic synthesis of behavior protocols for composable
  web-services.
\newblock In \emph{Proceedings of the 7th Joint Meeting of the {{European}}
  Software Engineering Conference and the {{ACM SIGSOFT}} Symposium on {{The}}
  Foundations of Software Engineering}, {{ESEC}}/{{FSE}} '09, pages 141--150,
  New York, NY, USA, August 2009. Association for Computing Machinery.
\newblock ISBN 978-1-60558-001-2.
\newblock \doi{10.1145/1595696.1595719}.
\newblock URL \url{https://dl.acm.org/doi/10.1145/1595696.1595719}.

\bibitem[Biermann and Feldman(1972)]{biermannSynthesisFiniteStateMachines1972}
A.~W. Biermann and J.~A. Feldman.
\newblock On the {{Synthesis}} of {{Finite-State Machines}} from {{Samples}} of
  {{Their Behavior}}.
\newblock \emph{IEEE Transactions on Computers}, C-21\penalty0 (6):\penalty0
  592--597, June 1972.
\newblock ISSN 1557-9956.
\newblock \doi{10.1109/TC.1972.5009015}.
\newblock URL \url{https://ieeexplore.ieee.org/abstract/document/5009015}.

\bibitem[Chen et~al.(2009)Chen, Farzan, Clarke, Tsay, and
  Wang]{chenLearningMinimalSeparating2009}
Yu-Fang Chen, Azadeh Farzan, Edmund~M. Clarke, Yih-Kuen Tsay, and Bow-Yaw Wang.
\newblock Learning {{Minimal Separating DFA}}'s for {{Compositional
  Verification}}.
\newblock In Stefan Kowalewski and Anna Philippou, editors, \emph{Tools and
  {{Algorithms}} for the {{Construction}} and {{Analysis}} of {{Systems}}},
  volume 5505, pages 31--45. Springer Berlin Heidelberg, Berlin, Heidelberg,
  2009.
\newblock ISBN 978-3-642-00767-5 978-3-642-00768-2.
\newblock \doi{10.1007/978-3-642-00768-2_3}.
\newblock URL \url{http://link.springer.com/10.1007/978-3-642-00768-2_3}.

\bibitem[Cho et~al.(2011)Cho, Babi{\'c}, Poosankam, Chen, Wu, and
  Song]{choMACEModelinferenceAssistedConcolic2011}
Chia~Yuan Cho, Domagoj Babi{\'c}, Pongsin Poosankam, Kevin~Zhijie Chen,
  Edward~XueJun Wu, and Dawn Song.
\newblock \{\vphantom\}{{MACE}}\vphantom\{\}:
  \{\vphantom\}{{Model-inference-Assisted}}\vphantom\{\} {{Concolic
  Exploration}} for {{Protocol}} and {{Vulnerability Discovery}}.
\newblock In \emph{20th {{USENIX Security Symposium}} ({{USENIX Security}}
  11)}, 2011.
\newblock URL
  \url{https://www.usenix.org/conference/usenix-security-11/mace-model-inference-assisted-concolic-exploration-protocol-and}.

\bibitem[Comparetti et~al.(2009)Comparetti, Wondracek, Kruegel, and
  Kirda]{comparettiProspexProtocolSpecification2009}
Paolo~Milani Comparetti, Gilbert Wondracek, Christopher Kruegel, and Engin
  Kirda.
\newblock Prospex: {{Protocol Specification Extraction}}.
\newblock In \emph{2009 30th {{IEEE Symposium}} on {{Security}} and
  {{Privacy}}}, pages 110--125, May 2009.
\newblock \doi{10.1109/SP.2009.14}.
\newblock URL \url{https://ieeexplore.ieee.org/abstract/document/5207640}.

\bibitem[Cook and Wolf(1998)]{cookDiscoveringModelsSoftware1998}
Jonathan~E. Cook and Alexander~L. Wolf.
\newblock Discovering models of software processes from event-based data.
\newblock \emph{ACM Transactions on Software Engineering and Methodology},
  7\penalty0 (3):\penalty0 215--249, July 1998.
\newblock ISSN 1049-331X.
\newblock \doi{10.1145/287000.287001}.
\newblock URL \url{https://dl.acm.org/doi/10.1145/287000.287001}.

\bibitem[Cui et~al.(2007)Cui, Kannan, and
  Wang]{cuiDiscovererAutomaticProtocol2007}
Weidong Cui, Jayanthkumar Kannan, and Helen~J Wang.
\newblock Discoverer: {{Automatic Protocol Reverse Engineering}}.
\newblock \emph{USENIX Security Symposium}, pages 1--14, 2007.

\bibitem[{de la Higuera}(2010)]{delahigueraGrammaticalInferenceLearning2010}
C.~{de la Higuera}.
\newblock \emph{Grammatical {{Inference}}: {{Learning Automata}} and
  {{Grammars}}}.
\newblock CUP, 2010.

\bibitem[Eltabakh et~al.(2006)Eltabakh, Eltarras, and
  Aref]{eltabakhSpacePartitioningTreesPostgreSQL2006}
M.Y. Eltabakh, R.~Eltarras, and W.G. Aref.
\newblock Space-{{Partitioning Trees}} in {{PostgreSQL}}: {{Realization}} and
  {{Performance}}.
\newblock In \emph{22nd {{International Conference}} on {{Data Engineering}}
  ({{ICDE}}'06)}, pages 100--100, April 2006.
\newblock \doi{10.1109/ICDE.2006.146}.

\bibitem[{Fiter{\u a}u-Bro{\c s}tean} et~al.(2017){Fiter{\u a}u-Bro{\c s}tean},
  Lenaerts, Poll, {de Ruiter}, Vaandrager, and
  Verleg]{fiterau-brosteanModelLearningModel2017}
Paul {Fiter{\u a}u-Bro{\c s}tean}, Toon Lenaerts, Erik Poll, Joeri {de Ruiter},
  Frits Vaandrager, and Patrick Verleg.
\newblock Model learning and model checking of {{SSH}} implementations.
\newblock In \emph{Proceedings of the 24th {{ACM SIGSOFT International SPIN
  Symposium}} on {{Model Checking}} of {{Software}}}, {{SPIN}} 2017, pages
  142--151, New York, NY, USA, July 2017. Association for Computing Machinery.
\newblock ISBN 978-1-4503-5077-8.
\newblock \doi{10.1145/3092282.3092289}.
\newblock URL \url{https://dl.acm.org/doi/10.1145/3092282.3092289}.

\bibitem[Giantamidis et~al.(2021)Giantamidis, Tripakis, and
  Basagiannis]{giantamidisLearningMooreMachines2021}
Georgios Giantamidis, Stavros Tripakis, and Stylianos Basagiannis.
\newblock Learning {{Moore}} machines from input--output traces.
\newblock \emph{International Journal on Software Tools for Technology
  Transfer}, 23\penalty0 (1):\penalty0 1--29, February 2021.
\newblock ISSN 1433-2787.
\newblock \doi{10.1007/s10009-019-00544-0}.
\newblock URL \url{https://doi.org/10.1007/s10009-019-00544-0}.

\bibitem[Grinchtein and
  Leucker(2006)]{grinchteinLearningFiniteStateMachines2006}
Olga Grinchtein and Martin Leucker.
\newblock Learning {{Finite-State Machines}} from {{Inexperienced Teachers}}.
\newblock In Yasubumi Sakakibara, Satoshi Kobayashi, Kengo Sato, Tetsuro
  Nishino, and Etsuji Tomita, editors, \emph{Grammatical {{Inference}}:
  {{Algorithms}} and {{Applications}}}, Lecture {{Notes}} in {{Computer
  Science}}, pages 344--345, Berlin, Heidelberg, 2006. Springer.
\newblock ISBN 978-3-540-45265-2.
\newblock \doi{10.1007/11872436_30}.

\bibitem[Grinchtein et~al.(2006)Grinchtein, Leucker, and
  Piterman]{grinchteinInferringNetworkInvariants2006}
Olga Grinchtein, Martin Leucker, and Nir Piterman.
\newblock Inferring {{Network Invariants Automatically}}.
\newblock In David Hutchison, Takeo Kanade, Josef Kittler, Jon~M. Kleinberg,
  Friedemann Mattern, John~C. Mitchell, Moni Naor, Oscar Nierstrasz,
  C.~Pandu~Rangan, Bernhard Steffen, Madhu Sudan, Demetri Terzopoulos, Dough
  Tygar, Moshe~Y. Vardi, Gerhard Weikum, Ulrich Furbach, and Natarajan Shankar,
  editors, \emph{Automated {{Reasoning}}}, volume 4130, pages 483--497.
  Springer Berlin Heidelberg, Berlin, Heidelberg, 2006.
\newblock ISBN 978-3-540-37187-8 978-3-540-37188-5.
\newblock \doi{10.1007/11814771_40}.
\newblock URL \url{http://link.springer.com/10.1007/11814771_40}.

\bibitem[Hagerer et~al.(2001)Hagerer, Margaria, Niese, Steffen, Brune, and
  Ide]{hagererEfficientRegressionTesting2001}
Andreas Hagerer, Tiziana Margaria, Oliver Niese, Bernhard Steffen, Georg Brune,
  and Hans-Dieter Ide.
\newblock Efficient {{Regression Testing}} of {{CTI-Systems}}.
\newblock \emph{Annual review of communication}, 55:\penalty0 1033--1040, 2001.

\bibitem[Lang et~al.(1998)Lang, Pearlmutter, and
  Price]{langResultsAbbadingoOne1998}
Kevin~J. Lang, Barak~A. Pearlmutter, and Rodney~A. Price.
\newblock Results of the {{Abbadingo}} one {{DFA}} learning competition and a
  new evidence-driven state merging algorithm.
\newblock In Vasant Honavar and Giora Slutzki, editors, \emph{Grammatical
  {{Inference}}}, Lecture {{Notes}} in {{Computer Science}}, pages 1--12,
  Berlin, Heidelberg, 1998. Springer.
\newblock ISBN 978-3-540-68707-8.
\newblock \doi{10.1007/BFb0054059}.

\bibitem[Leucker and Neider(2012)]{leuckerLearningMinimalDeterministic2012}
Martin Leucker and Daniel Neider.
\newblock Learning {{Minimal Deterministic Automata}} from {{Inexperienced
  Teachers}}.
\newblock In Tiziana Margaria and Bernhard Steffen, editors, \emph{Leveraging
  {{Applications}} of {{Formal Methods}}, {{Verification}} and {{Validation}}.
  {{Technologies}} for {{Mastering Change}}}, Lecture {{Notes}} in {{Computer
  Science}}, pages 524--538, Berlin, Heidelberg, 2012. Springer.
\newblock ISBN 978-3-642-34026-0.
\newblock \doi{10.1007/978-3-642-34026-0_39}.

\bibitem[Moeller et~al.(2023)Moeller, Wiener, {Solko-Breslin}, Koch, Foster,
  and Silva]{moellerAutomataLearningIncomplete2023}
Mark Moeller, Thomas Wiener, Alaia {Solko-Breslin}, Caleb Koch, Nate Foster,
  and Alexandra Silva.
\newblock Automata {{Learning}} with an {{Incomplete Teacher}}.
\newblock In Karim Ali and Guido Salvaneschi, editors, \emph{37th {{European
  Conference}} on {{Object-Oriented Programming}} ({{ECOOP}} 2023)}, volume 263
  of \emph{Leibniz {{International Proceedings}} in {{Informatics}}
  ({{LIPIcs}})}, pages 21:1--21:30, Dagstuhl, Germany, 2023. Schloss Dagstuhl
  -- Leibniz-Zentrum f{\"u}r Informatik.
\newblock ISBN 978-3-95977-281-5.
\newblock \doi{10.4230/LIPIcs.ECOOP.2023.21}.
\newblock URL \url{https://drops.dagstuhl.de/opus/volltexte/2023/18214}.

\bibitem[Oncina and Garcia(1992)]{oncinaIdentifyingRegularLanguages1992}
Jos'e Oncina and Pedro Garcia.
\newblock Identifying regular languages in polynomial time.
\newblock \emph{Advances in Structural and Syntactic Pattern Recognition},
  pages 99--108, 1992.
\newblock URL
  \url{https://www.worldscientific.com/doi/abs/10.1142/9789812797919_0007}.

\bibitem[Schmidt and Kramer(2014)]{schmidtOnlineInductionProbabilistic2014}
Jana Schmidt and Stefan Kramer.
\newblock Online {{Induction}} of {{Probabilistic Real-Time Automata}}.
\newblock \emph{Journal of Computer Science and Technology}, 29\penalty0
  (3):\penalty0 345--360, May 2014.
\newblock ISSN 1860-4749.
\newblock \doi{10.1007/s11390-014-1435-8}.
\newblock URL \url{https://doi.org/10.1007/s11390-014-1435-8}.

\bibitem[Vaandrager et~al.(2022{\natexlab{a}})Vaandrager, Garhewal, Rot, and
  Wi{\ss}mann]{vaandragerNewApproachActive2022}
Frits Vaandrager, Bharat Garhewal, Jurriaan Rot, and Thorsten Wi{\ss}mann.
\newblock A {{New Approach}} for {{Active Automata Learning Based}} on
  {{Apartness}}.
\newblock In Dana Fisman and Grigore Rosu, editors, \emph{Tools and
  {{Algorithms}} for the {{Construction}} and {{Analysis}} of {{Systems}}},
  Lecture {{Notes}} in {{Computer Science}}, pages 223--243, Cham,
  2022{\natexlab{a}}. Springer International Publishing.
\newblock ISBN 978-3-030-99524-9.
\newblock \doi{10.1007/978-3-030-99524-9_12}.

\bibitem[Vaandrager et~al.(2022{\natexlab{b}})Vaandrager, Garhewal, Rot, and
  Wi{\ss}mann]{vaandragerNewApproachActive2022a}
Frits Vaandrager, Bharat Garhewal, Jurriaan Rot, and Thorsten Wi{\ss}mann.
\newblock A {{New Approach}} for {{Active Automata Learning Based}} on
  {{Apartness}}, January 2022{\natexlab{b}}.
\newblock URL \url{http://arxiv.org/abs/2107.05419}.

\bibitem[Verwer and Hammerschmidt(2022)]{verwerFlexFringeModelingSoftware2022}
Sicco Verwer and Christian Hammerschmidt.
\newblock {{FlexFringe}}: {{Modeling Software Behavior}} by {{Learning
  Probabilistic Automata}}, 2022.
\newblock URL \url{http://arxiv.org/abs/2203.16331}.

\bibitem[Verwer and Hammerschmidt(2017)]{verwerFlexfringePassiveAutomaton2017}
Sicco Verwer and Christian~A. Hammerschmidt.
\newblock Flexfringe: {{A Passive Automaton Learning Package}}.
\newblock In \emph{2017 {{IEEE International Conference}} on {{Software
  Maintenance}} and {{Evolution}} ({{ICSME}})}, pages 638--642, September 2017.
\newblock \doi{10.1109/ICSME.2017.58}.
\newblock URL \url{https://ieeexplore.ieee.org/abstract/document/8094471}.

\end{thebibliography}
\let\clearpage\relax
\end{document}